**The $J_{eff}$=1/2 antiferromagnet $Sr_2IrO_4$: A golden avenue toward new physics and functions**

Chengliang Lu* and Jun-Ming Liu*

Dr. C. L. Lu

School of Physics & Wuhan National High Magnetic Field Center,

Huazhong University of Science and Technology, Wuhan 430074, China

Email: cllu@hust.edu.cn

Prof. J. –M. Liu

1. Laboratory of Solid State Microstructures and Innovation Center of Advanced Microstructures, Nanjing University, Nanjing 210093, China

2. Institute for Advanced Materials, South China Normal University, Guangzhou 510006, China

Email: liujm@nju.edu.cn

**Abstract**

Iridates have been providing a fertile ground for studying emergent phases of matter that arise from delicate interplay of various fundamental interactions with approximate energy scale. Among these highly focused quantum materials, perovskite $Sr_2IrO_4$ belonging to the Ruddlesden-Popper series stands out and has been intensively addressed in the last decade, since it hosts a novel $J_{eff} = 1/2$ state which is a profound manifestation of strong spin-orbit coupling. Moreover, the $J_{eff} = 1/2$ state represents a rare example of iridates that has been better understood both theoretically and experimentally. In this progress report, we take $Sr_2IrO_4$ as an example to overview the recent advances of the $J_{eff} = 1/2$ state in two aspects: materials fundamentals and functionality potentials. In the fundamentals part, we first illustrate basic issues for the layered canted antiferromagnetic order of the $J_{eff} = 1/2$ magnetic moments in $Sr_2IrO_4$, and then review the progress of the antiferromagnetic order modulation through diverse routes. Subsequently, for the functionality potentials, fascinating properties such as atomic-scale giant magnetoresistance, anisotropic magnetoresistance, and nonvolatile memory, will be addressed. This report will be concluded with our prospected remarks and outlooks.

**1. Introduction**

Electron-electron interaction ($U$) and spin-orbit coupling (SOC, measured by strength factor $\lambda$) are two critical fundamental ingredients that determine the electronic properties and functionalities of a quantum material [1]. The correlated electron materials have been greatly emphasized in 3$d$ transition metal oxides where diverse magnetic textures, metal-insulator transitions, unconventional superconductivity etc. were found. However, the SOC of 3$d$ elements is weak and usually treated as perturbations to the electron correlations in materials ($\lambda$ ~0.01 eV and $U$~5 eV for 3d ions). One exception could be multiferroics with non-collinear magnetic configuration where SOC doesn't have to be large but still plays a dominant role in generating ferroelectricity [2, 3]. It is known that SOC is effectively increased with atomic number, and consequently, emergent non-trivial quantum states can be induced. Topological insulators and semimetals are profound manifestations of the enhanced SOC [4, 5]. In these extremely hot topics, the Hubbard repulsion $U$ is usually small (if not all), giving rise to a barrier to develop intrinsic magnetism [1, 6].

Iridates certainly are another group of materials that show remarkably enhanced SOC ($\lambda$ ~0.5 eV). Importantly and also differently, iridates in the meanwhile possess a moderate $U$ (~ 2.0 eV), which provides a novel platform for cooperative effects of $U$ and SOC [7, 8], and this cooperation is unique, characterized by the fact that various exchange energy scales, Hubbard repulsion $U$, SOC, and other competing interactions are comparable with each other. Theoretical investigations revealed that the delicate interplay of these fundamental interactions can cause a large array of novel electronic states, including the $J_{\text{eff}}$ = 1/2 Mott state [9, 10], Weyl semimetals with Fermi arcs [5], correlated topological insulator [1], Kitaev spin liquid [11], excitonic magnetism of pentavalent $Ir^{5+}$ (5$d^4$) [12], etc. Probably due to the rapid developing nature of this field, there are just a few of these proposals that have been demonstrated experimentally, resulting in a huge gap between theoretical predictions and experimental findings. Among these, the $J_{eff} = 1/2$ Mott state represents a rare but unique example that has been better understood both theoretically and experimentally and mostly addressed in the past decade.

Since the 5$d$ orbitals of Ir are more extended in comparison with 3$d$ elements, a common perception assumes that iridates would be more metallic and less magnetic than the 3$d$ compounds. However, since 1990s and early 2000s, a couple of magnetic and insulating iridates have been successfully synthesized [13-15], and the observed results are distinctly different from our current consensus. This striking conflict was partially solved by identification of the $J_{eff}=1/2$ picture in Sr$_2$IrO$_4$ in 2008 [9, 10]. As shown in **Figure 1**, because of crystal field effect, the 5$d$ orbitals of Ir are split into the four-fold degenerate $e_g$ orbitals and six-fold $t_{2g}$ orbitals, and the $e_g$ state takes a much higher energy than the $t_{2g}$ one. In this sense, the five electrons (Ir$^{4+}$) reside in the $t_{2g}$ orbitals with an effective orbital angular momentum $L = 1$. In the presence of strong SOC which acts within the $t_{2g}$ manifold as -$\lambda \bm{L} \cdot \bm{S}$, where L is the effective angular momentum and S is the spin moment, the $t_{2g}$ band is split into an effective $J_{eff} = 1/2$ doublet and an effective $J_{eff} = 3/2$ quartet. As a result, the $J_{eff} = 3/2$ band is fully occupied with four electrons due to its lower energy, and the energetically higher $J_{eff} = 1/2$ band is half-filled with one remaining electron. The $J_{eff} = 1/2$ band is thin enough, and a moderate $U$ can thus open a charge-gap ranging from ~ 0.1 eV to ~ 0.5 eV [16].

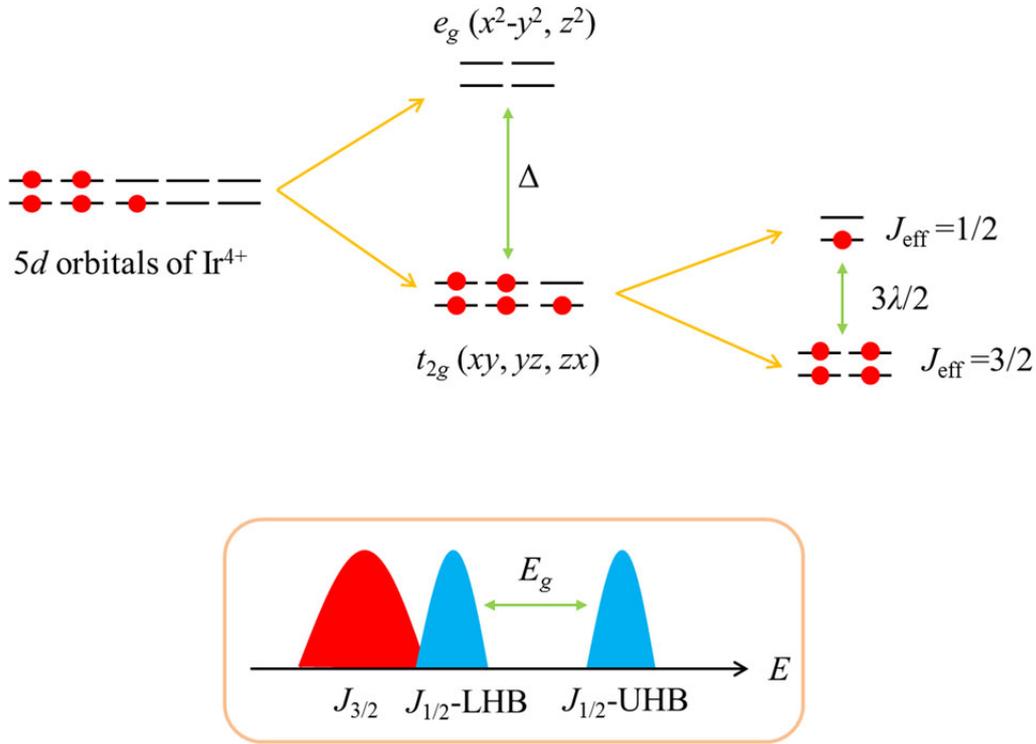

Figure 1. A sketch of the $J_{eff} = 1/2$ model where the 5d orbitals of $Ir^{4+}$ are split by crystal field and spin-orbit coupling (which assumes $\boldsymbol{L}\cdot\boldsymbol{S}$ coupling), and finally a charge-gap is opened due to the Coulomb interaction. LHB and UHB represent the lower Hubbard band and upper Hubbard band, respectively.

The $J_{eff} = 1/2$ model has been successfully utilized to explain the insulating nature of many iridates [9, 15, 17-19], and it works quite well although the ideal $J_{eff} = 1/2$ state is only expected for a perfect cubic symmetry. For instance, layered perovskites $Sr_2IrO_4$ and $Sr_3Ir_2O_7$ (Ruddlesden Popper compounds, $Sr_{n+1}Ir_nO_{3n+1}$, $n = 1$ and 2) have been extensively discussed within the $J_{eff}=1/2$ scenario, although theoretical calculations have found that the both show certain deviation from the ideal $J_{eff}=1/2$ situation related to the tetragonal distortion [20]. Nevertheless, a deviation from a cubic crystal environment still exhibits non-negligible influence to the electronic properties, especially to magnetism, which has been a central issue within the domain of iridates researches.

Apart from the fundamental aspects, the $J_{eff} = 1/2$ state also shows tantalizing functionalities, which has been reported typically in the leading iridate $Sr_2IrO_4$ hosting a canted antiferromagnetic (AFM) structure [21-28]. Since $Sr_2IrO_4$ and $La_2CuO_4$ show couple of similarities in terms of structure and magnetic configuration, possible unconventional superconducting behaviors (even *p*-wave superconductivity and multipole superconductivity related to an odd parity magnetic hidden order) was frequently discussed [29-32]. This however remains elusive in experiments, and the two signature features of superconductivity, i.e. zero resistance and diamagnetism, are still missed. Another noteworthy phenomenon is the significantly large magnetoresistance (MR) in $Sr_2IrO_4$ with a robust AFM order, unveiling tantalizing functionalities of the $J_{eff} = 1/2$ state toward the antiferromagnetronics (AFMtronics, referred to AFM spintronics in some occasions) [21-28, 33]. These phenomena have underscored the uniqueness of iridates with the $J_{eff} = 1/2$ state as promising and unique functional materials.

Surely, the family of iridates as a new group of quantum materials possess many more fascinating properties beyond the $J_{eff} = 1/2$ state, and there are indeed several review articles which cover this topic partially or to some extent [1, 7, 8, 34]. In this progress report, we take $Sr_2IrO_4$ as representative material to address recent advances of the $J_{eff} = 1/2$ state, from fundamentals to functionalities, while no such framework has been touched with sufficient depth and breadth. The earlier reviews mainly covered the basic properties of iridates including $Sr_2IrO_4$ [8, 35], here we focus on the novel functionalities associated with the $J_{eff} = 1/2$ state, including the SOC twisted antiferromagnetism and AFMtronics effects. This report is thus organized as the following. In Section 2, we illustrate the fundamental aspects of the canted AFM order with $Sr_2IrO_4$. In Section 3, we overview the recent advances in terms of the SOC twisted antiferromagnetism. In Section 4, we describe the distinctly large MR in $Sr_2IrO_4$, which underscores the uniqueness of $Sr_2IrO_4$ as one of AFM electronic materials. Remarkable and controllable anisotropic magnetoresistance (AMR) and nonvolatile memory effect in $Sr_2IrO_4$ are further discussed in Section 5. At the end of this progress report, concluding remarks and outlooks, likely with our personal bias, will be presented. Given the rapid evolving nature of this field, we certainly don't have any attempt to cover everything in this short article, and may miss some important results in this field.

## 2. $J_{eff}$ = 1/2 antiferromagnetism in $Sr_2IrO_4$

$Sr_2IrO_4$ is a single-layer perovskite, belonging to the Ruddlesden Popper series $Sr_{n+1}Ir_nO_{3n+1}$ with $n = 1$ [36]. It was reported to crystalline in a tetragonal structure with space group $I4_1/acd$. This assignment has been challenged, and a reduced structural symmetry with space group $I4_1/a$ has been proposed [37, 38]. The $IrO_6$ octahedra rotate about the $c$-axis by an angle $\alpha \sim 11°$, leading to an expanded unit cell with $a = b = 5.4846$ Å and $c = 25.804$ Å [37, 39]. The tetragonal distortion makes $Sr_2IrO_4$ deviate from the ideal $J_{eff} = 1/2$ situation which is only expected in a cubic crystal environment. Theoretical calculations found that the deviation is only ~ 3 % [20], confirmed by further non-resonant magnetic X-ray scattering (NRMXS) on $Sr_2IrO_4$ [40]. In the meanwhile, the NRMXS experiments revealed a significantly enhanced orbital component of the $J_{eff} = 1/2$ magnetic moment (hereafter called the pseudospin), which is much larger than the ideal $J_{eff} = 1/2$ state [40]. Possible hybridization of $e_g$ and $t_{2g}$ states, and the weak Mott character were suggested as additional physical sources beside the elongated octahedral along the $c$-axis. According to the $J_{eff} = 1/2$ model, the pseudospin with equal contributions from the three $t_{2g}$ orbitals is isotropic and SU(2) invariant, and carries a magnetic moment of 1.0 $\mu_B$/Ir [20]. However, for $Sr_2IrO_4$, a much smaller magnetic moment ~0.3 $\mu_B$/Ir due to the extended orbital has been identified, and the SU(2) invariance is broken because of the tetragonal distortion [37, 41].

The pseudospins in $Sr_2IrO_4$, entangling both spin and orbital momenta due to the strong SOC at Ir-site, are aligned into an AFM lattice below the Néel temperature $T_N$ ~ 240 K [13, 24, 25, 28], as illustrated in **Figure 2**. The absence of anomaly at $T_N$ in the electrical transport data is in agreement with the scenario of $J_{eff}$ = 1/2 Mott state. All the pseudospins lie in the *ab*-plane, and show uniform deviation of ~13° from the *a*-axis. The canting is within the *ab*-plane, and there is no canting along the *c*-axis. Various techniques, including neutron scattering and resonant X-ray scattering (RXS), demonstrated that the pseudospin canting rigidly tracks the rotation of $IrO_6$ octahedra, resulting in the well-known locking relation: $\alpha \sim \phi$ [37, 39, 41]. This locking effect between the pseudospin canting and lattice distortion was reproduced theoretically by Jackeli *et al.* [42]. Because of the canting, net magnetic moments appear alternatively in each $IrO_2$ planes, and are coupled antiferromagnetically along the *c*-axis. In this sense, $Sr_2IrO_4$ is fully compensated at the ground state without showing macroscopic magnetization. A magnetic field $H$ larger than the critical value $H_{flop}$ can cause a flop transition, and the net moments of $IrO_2$ layers are then ferromagnetically aligned along the *c*-axis, resulting in a weak ferromagnetic (FM) phase [43, 44], as sketched in Fig. 2. This critical field $H_{flop}$ is fortunately not large (~ 0.1 T), and it together with the flop transition makes $Sr_2IrO_4$ very unique. Consequently a number of emergent phenomena have been observed.

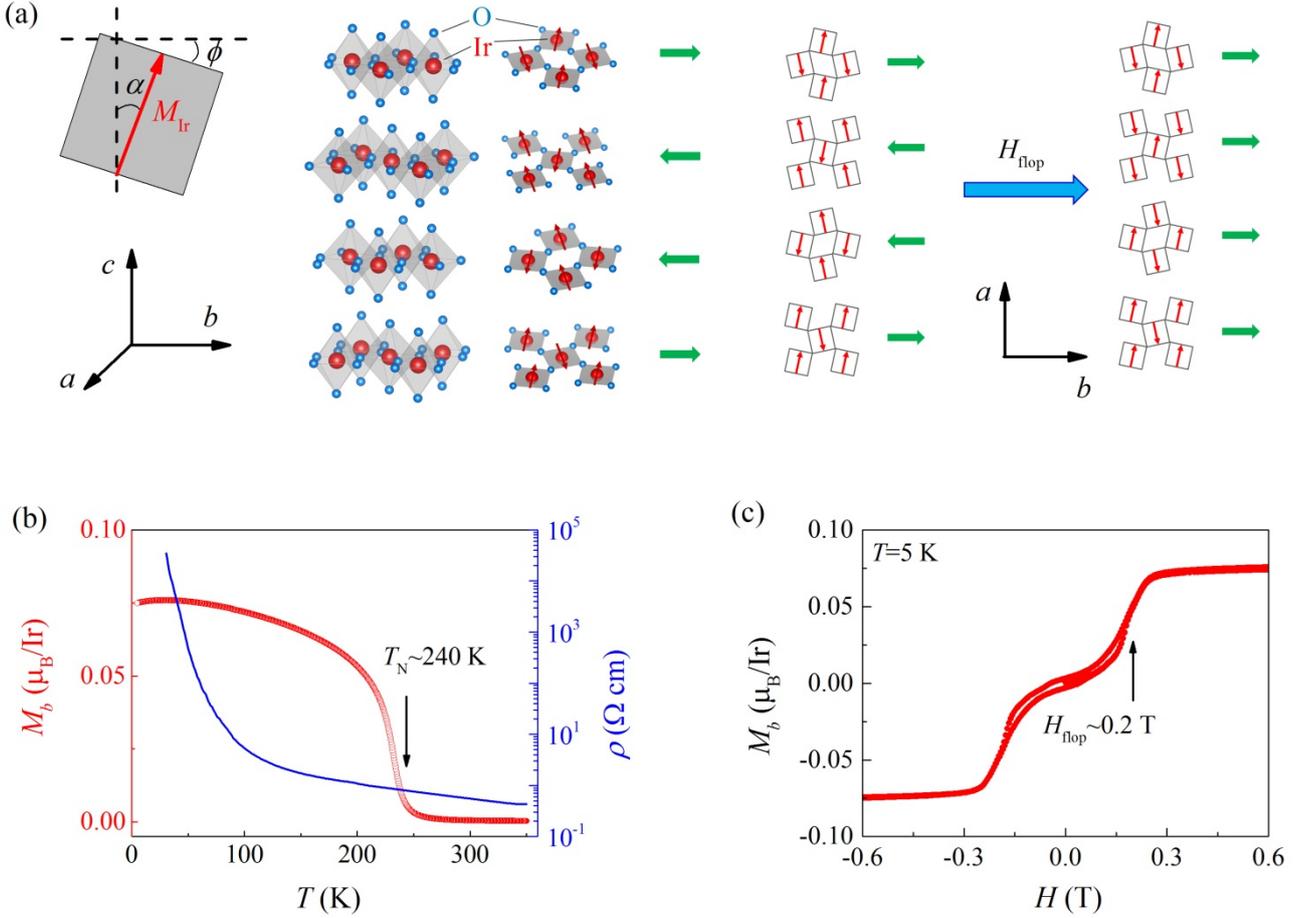

Figure 2. (a) A sketch of crystal and magnetic structures of $Sr_2IrO_4$, and the so-called pseudospin flop transition driven by magnetic field $H$ larger than the critical field $H_{flop}$. The magnetic moments of Ir are indicated by red arrows, and the net moments $\mu_{net}$ of $IrO_2$ layers are indicated by green arrows. (b) The temperature dependence of $b$-axis magnetization $M_b$ and in-plane resistivity $\rho$, where $T_N \sim 240$ K is indicated. (c) The step-anomaly in the $M_b(H)$ curve due to the flop transition.

Benefiting from recent technological advances in resonant X-ray scattering, direct probing of the Ir pseudospins is allowed, noting that Ir has large neutron absorption ratio, which challenges related characterizations using neutron scattering technique [37]. Extensive efforts in the last several years revealed that the AFM order is mainly stabilized by the strong nearest-neighbor (NN) intralayer AFM interaction of ~ 0.1 eV [45-47]. Regarding the interlayer coupling, it was found to be as small as ~ 1.0 μeV, $10^5$ times weaker than the intra-plane AFM coupling [47]. Such a large magnetic anisotropy allows the so-called flop transition, also known as the AFM to weak FM (AFM-wFM) transition, as demonstrated with RXS [43]. The AFM order of $Sr_2IrO_4$ looks fairly robust and the AFM fluctuation with long-rang in-plane correlation was found to persist at even ~ 20 K above $T_N$, which can be described by the two-dimensional $S = 1/2$ quantum Heisenberg model [47]. Another work addressing on critical behavior of $Sr_2IrO_4$ revealed that the critical scattering can be followed out to a much higher temperature ($T_N$+73 K), and XY anisotropy is important for accounting for the pseudospin interactions [48]. High magnetic field experiments revealed that the AFM order doesn't show any trace of breakdown up to $H \sim 60$ T [25]. The weak FM phase was found to be quenched at hydrostatic pressure $P \sim 20$ GPa, which is probably due to the reorientation of pseudospins from the $ab$-plane to the $c$-axis rather than the AFM order collapse [49].

While the NN interaction in developing the long range AFM order has been emphasized, some further next-order interactions with smaller energy scales seem to be non-negligible to the low energy magnetic properties of $Sr_2IrO_4$. By performing theoretical calculations and various microscopic measurements, Porras *et al.* revealed that pseudospin-lattice coupling is crucial to understand the static magnetism and low-energy pseudospin dynamics in $Sr_2IrO_4$ [44]. This was discussed and confirmed in the meantime by pure theoretical considerations [50]. After incorporating the term of pseudospin-lattice coupling into the effective Hamiltonian, one may unravel many puzzles of $Sr_2IrO_4$ such as in-plane magnetic anisotropy and reduced structural symmetry. It is worthy to note that the in-plane magnetic anisotropy has been a long-term controversial issue, which is now clarified by improving crystal quality [25, 43, 51, 52]. This issue will be further discussed in the following, as it hosts emergent functionalities, i.e. as an AFMtronics ingredient.

## 3. Tuning the antiferromagnetism

*3.1 Tuning via carrier doping*

In parallel to continuous interest in the $J_{eff}$ = 1/2 antiferromagnetism, enthusiastic endeavors have been given also to carrier doping in $Sr_2IrO_4$, including electron- and hole-dopings, owing to not only the predicted unconventional superconductivity. Indeed, a bunch of features such as Fermi arcs, pseudogaps, and inhomogeneous electronic order have been observed [53-57], which are highly parallel to those seen in cuprates. However, the two signatures of superconductivity, i.e. zero resistivity and diamagnetism, have not yet been observed. Recently, a hidden broken-symmetry phase was identified in $Sr_2IrO_4$, and Rh-doping (hole-doping) can enlarge the region of hidden phase in the phase diagram [38, 58]. Probably driving this hidden phase to a quantum critical point by heavier doping would be an approach to the missed superconductivity. Since electron-doping behaves more efficient than hole-doping in generating metallicity in $Sr_2IrO_4$, an interested question is how the hidden order would evolve upon electron-doping, which is still unexplored.

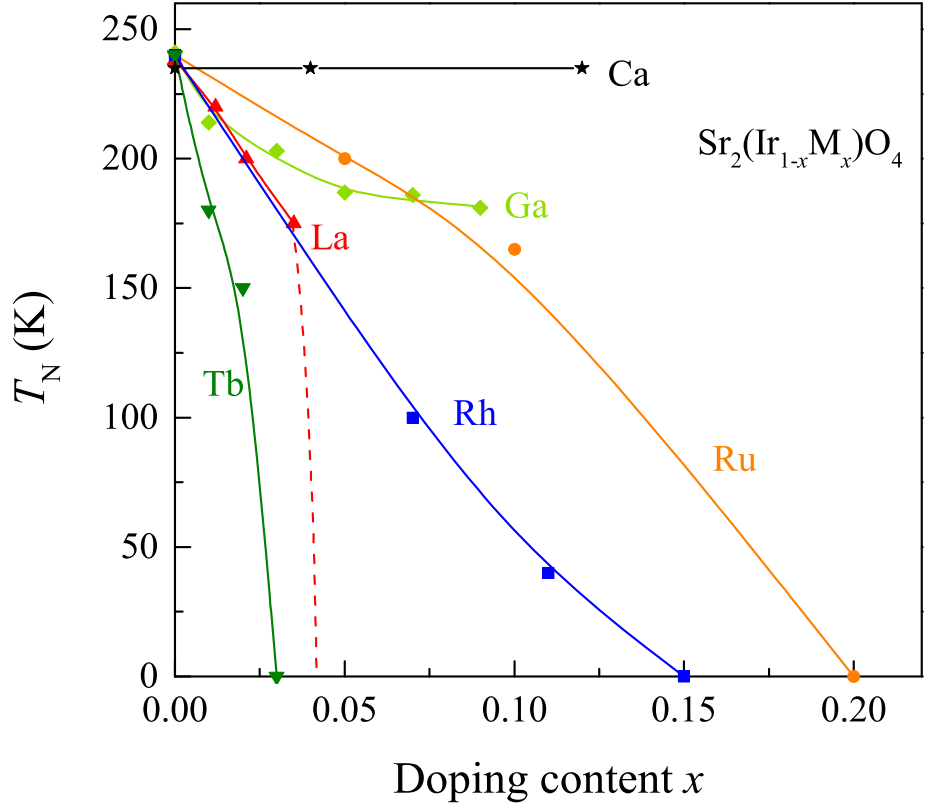

Figure 3. The Néel point $T_N$ as a function of doping content $x$ for various dopants at the Ir-site as labeled. (Data were collected from References [24], [63], [67], [69], [74]).

While the proposed superconductivity warrants further verification in experiments, the studies that have been made so far focusing on this topic have revealed interesting phenomena on the magnetism. The Néel temperature $T_N$ as a function of doping level for various dopants is shown in **Figure 3**. First, electron-doping and hole-doping in $Sr_2IrO_4$ show highly asymmetric effect on the AFM phase. For the electron-doping, i.e. La-substitution at Sr-site, the long-range AFM order is rapidly lost at a doping level of only ~ 3 %, but the short-range correlated orders persist well into the metallic region [28, 59-62]. Further increasing the La-content slightly can cause a unidirectional spin density wave state [63]. On the contrary, the long-range AFM order exhibits much more robustness against the hole-doping. For instance, in $Sr_2Ir_{1-x}Ga_xO_4$, the long range AFM phase can be well preserved, and $T_N$ just shows modest decrease from ~ 240 K to ~ 180 K with increasing $x$ up to

~ 9 % [24]. While Rh was demonstrated as hole-dopant, the AFM phase transition can still be detected even at a doping level as high as $x \sim 11$ % [64, 65]. The slight K-substitution of Sr can break the Mott state easily akin to electron doping effect, but it doesn't show any apparent impact onto the AMF order [28]. Oxygen deficiency in materials is known to be equivalent to electron-doping. In $Sr_2IrO_4$, generating oxygen vacancies can break the Mott state quickly, but the canted AFM order remains nicely [66]. This is similar to the K-doping effect (hole doping), but distinctly different from the case of La-doping (electron doping) which destroys both the charge-gap and the long-range AFM order simultaneously.

Second, the Ir-site substitution in $Sr_2IrO_4$ looks quite complicated. The 10% Mn-doping at Ir-site can flop the pseudospins from the basal plane to the *c*-axis, but doesn't show much effect on the magnetization and electric transport [67]. Similar pseudospin reorientation has also been observed in $Sr_2Ir_{1-x}Ru_xO_4$ [68]. According to the model often used for magnetism in layered iridates [42, 69], a modified interlayer coupling may be responsible for such a flop transition from the *ab*-plane to the *c*-axis. A slight isovalent substitution of Ir with Tb (~ 3 %) causes the long-range AFM order to collapse, which was interpreted by a compass impurity model [70, 71]. These works have shown the important role of magnetic cation doping at Ir-site. It has been found that the Ga- and Rh-doping can also trigger the AFM-wFM transition [24, 72], resembling the case driven by magnetic field *H*. Theoretical calculations suggest that a hole-doping may be important for the in-plane flopping transition [73]. However, similar effect was recently reported in $Sr_2Ir_{1-x}Sn_xO_4$ where Sn is expected to take a tetravalent state as $Ir^{4+}$ [74].

Third, substitution of Sr with isovalent cation such as Ca or Ba is expected to cause lattice distortion only, without implanting charge carriers and additional magnetic perturbations. The Ca-doping tends to shrink the lattice, while the Ba doping prefers to tune the Ir-O-Ir bond angle [75], which can be increased by ~ 1° due to 4% Ba-doping, and 15% Ca-doping can shrink the cell volume by ~ 1 %. It is noted that the electronic band structure of $Sr_2IrO_4$ is correlated with the Ir-O-Ir bond angle, and indeed a sudden drop of the in-plane resistance at low temperature has been observed for the sample with 2% Ba. Regarding the magnetic properties, no apparent change of the Néel temperature $T_N$ with the Ca- or Ba-doping can be seen [75, 76], as shown in Fig. 3. Nevertheless, the Ca-doping can evidently reduce the ratio of the out-of-plane magnetization to the in-plane one, $M_a/M_c$, indicating the suppressed magnetic anisotropy.

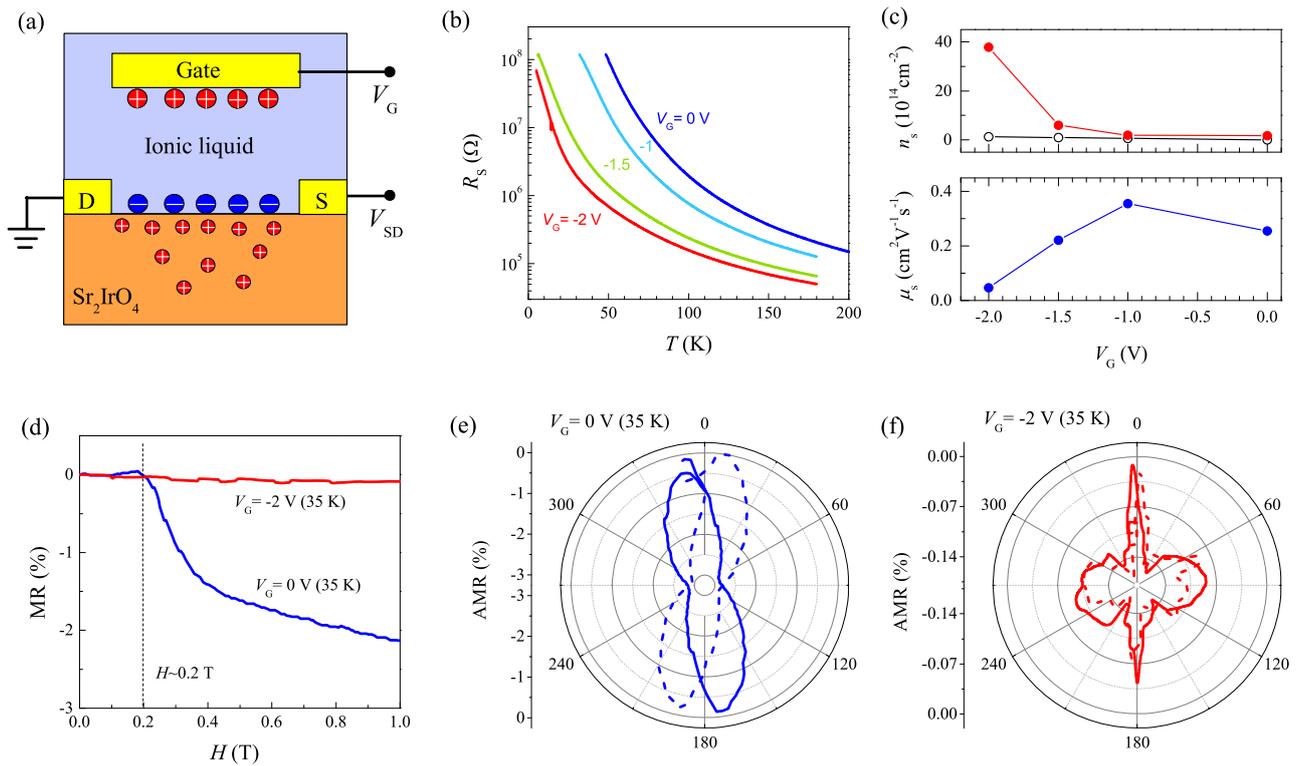

Figure 4. (a) A sketch of the field effect transistor with $Sr_2IrO_4$ where the gate layer is an ionic liquid. (b) The temperature dependence of resistance $R_S$ under various gating voltage $V_G$. (c) The charge carrier density and mobility as a function of gating voltage $V_G$. (d) Magnetoresistance under $V_G$ =0 and − 2.0 V at $T$ = 35 K. (e) and (f) present the anisotropic magnetoresistance (AMR) measured under $V_G$ = 0 V and - 2.0 V at $T$ = 35 K, respectively. The dashed and solid lines in (f)

represent the AMR trace and retrace. Reproduced with permission from Ref. [33].

It should be mentioned that one of the major difficulties for chemical substitution in $Sr_2IrO_4$ is the limited tolerance for doping, which hinders the role of carrier density modulation. Alternative approaches include the field effect transistor (FET) scheme which is one of the most elegant ways to inject carriers into materials without generating parasitic effects like quenching disorder. This scheme with a modified structure, i.e. electric-double-layer transistor (EDLT) with electrolyte as the gating layer, has been recognized as a very powerful method to tune the carrier density [77]. Recently, the EDLT with $Sr_2IrO_4$ as the channel layer was successfully fabricated to tune the $J_{eff} = 1/2$ state [33, 78], as shown in **Figure 4**. Indeed, the carrier density in the $Sr_2IrO_4$ layer can be drastically enhanced up to $\sim 4 \times 10^{15}$ /cm$^2$, about two orders of magnitude higher than that with the non-gated sample [33]. Unfortunately, no superconductivity was yet detected and the transport data show insulating behavior. It should be mentioned that this carrier density is already higher than that in a lot of superconductors [79]. On the other hand, the enhanced carrier density did evidence significant impact on magnetotransport such as standard MR and AMR, indicating the gating-modified magnetic properties in $Sr_2IrO_4$ [33, 78]. Furthermore, a linear relationship between conductivity and channel thickness implies a bulk gating effect of the EDLT structure.

*3.2 Tuning via strain*

Besides the chemical doping, $Sr_2IrO_4$ offers another roadmap of tunability, considering the locking effect $\alpha \sim \phi$, which suggests another idea to engineer the AFM order of epitaxial thin films via lattice strain. This idea was initially examined theoretically, and indeed remarkable strain effects on magnetism were revealed. Upon a strain, the ground state can be shifted toward or from the $SU(2)$ point, and the canting angle of pseudospins and the $c/a$ ratio of $IrO_6$ octahedra are all linearly proportional to strain [20, 42, 80]. As shown in **Figure 5**, when parameters $c/a$ and $\alpha$ are small, an *ab*-plane collinear AFM state appears. Increasing the $c/a$ ratio would rotate the pseudospins from the *ab*-plane to the *c*-axis, yet keeping the collinear AFM configuration. However, upon the increased rotation of $IrO_6$ octahedra, the pseudospins are expected to be clearly canted because of the locking effect $\alpha \sim \phi$. The electronic structure is predicted to be sensitive to strain [81], evidenced by the enhanced electron correlation and spin-orbit coupling with increasing strain, while the response of electronic structure to strain is highly correlated with the direction of magnetic moment. The density-functional calculations do suggest the strong strain-dependence of transport properties for hole-doped $Sr_2IrO_4$ [82].

Experimentally, it was found that the Néel point $T_N$ depends remarkably on strain, and $T_N$ can be drastically enhanced (reduced) by tensile (compressive) strain in $Sr_2IrO_4$ epitaxial films [83]. In comparison with the bulk counterpart, $T_N$ of the (001) $Sr_2IrO_4$/$SrTiO_3$ thin films with a ~ 0.5 % tensile strain is enhanced by ~ 30 K. The underlying physics can be ascribed to the renormalization of interlayer and intralayer interactions due to the structural engineering by strain. However, this observation was questioned by other reports on the same system where the observed $T_N$ is even lower than that of the bulk counterpart [26, 33]. Similar inconsistence can be found for the electrical transport data. In fact, for bulk $Sr_2IrO_4$, a tiny La-doping is sufficient to drive the insulating to metallic transition, and this transition is understandable if one considers the small band gap of the $J_{eff} = 1/2$ state. However, the insulating ground state in epitaxial thin films seems to be highly robust even at a much higher doping level [78, 84]. Another example to show such inconsistence is $(Sr_{1-x}Ba_x)_2IrO_4$ which in the bulk form shows strong metallic behavior at $x = 0.02$. The intriguing insulating transport remains robust even at very high Ba-content ($x \sim 37.5\ \%$) [75, 85].

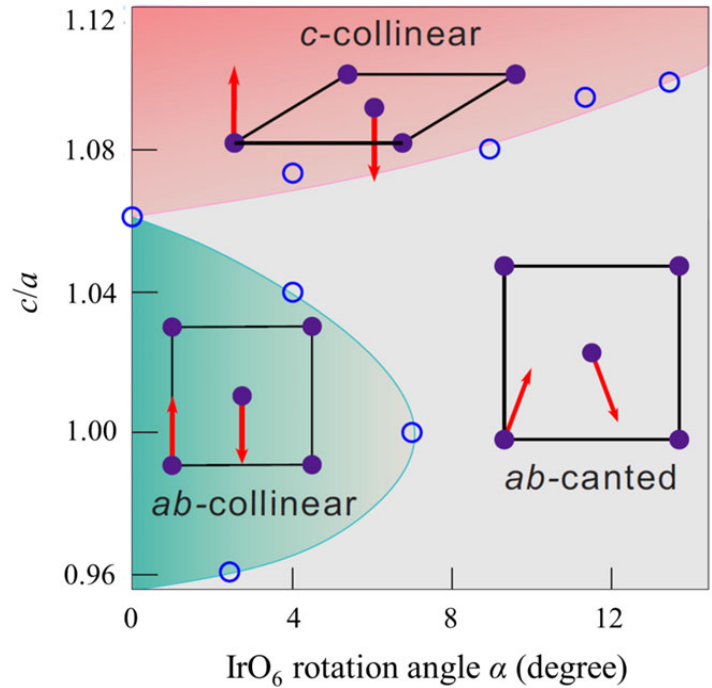

Figure 5. The calculated magnetic phase diagram of $Sr_2IrO_4$, in which various magnetic structures can be obtained, including the *ab*-collinear, *c*-collinear, and *ab*-canted phases. Reproduced with permission from Ref. [80].

It should be mentioned that pure $Sr_2IrO_4$ phase can only be stabilized within a relatively narrow growth window and the electronic properties show high sensitivity to chemical environment (i.e. defects and interfaces) [86-89]. In fact, even for high quality $Sr_2IrO_4$ epitaxial films, crossover behavior in terms of conduction mechanism upon varying thickness and temperature was observed, as shown in **Figure 6** [90]. Nevertheless, one has to admit that high doping concentration is allowed in thin film samples, a favored advantage that is not accessible in the bulk. In agreement with theoretical calculations, the strain modified electron correlation and electronic band structure were experimentally revealed in $Sr_2IrO_4$ thin films grown on various substrates [91-94], while such agreement does not happen for the bulk crystals.

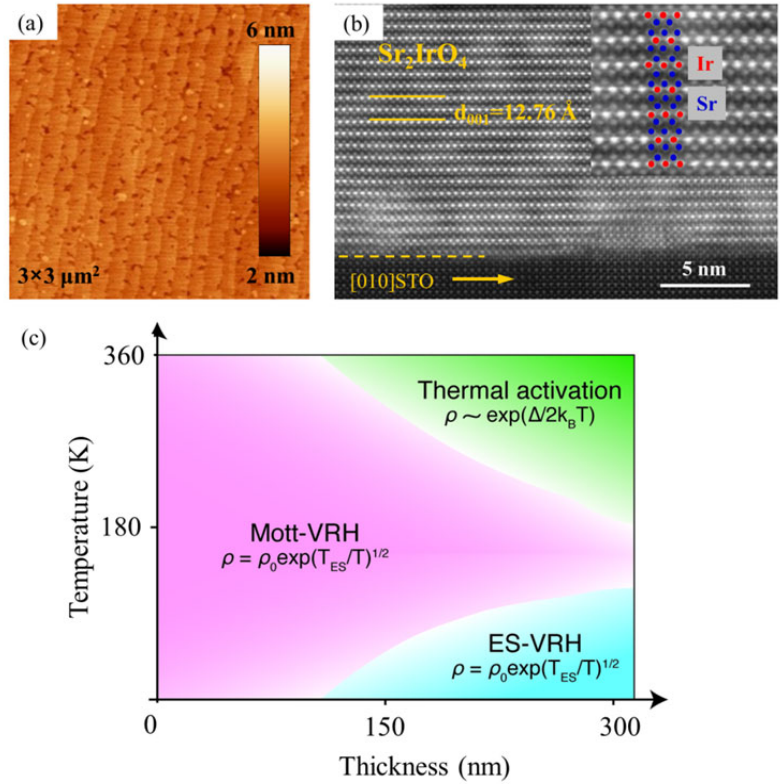

Figure 6. (a) Surface morphology with clear terrace surface structure, and (b) scanning transmission electron microscopy image showing nice SrO and $IrO_2$ planes stacking for a $Sr_2IrO_4$ thin film of 18 nm in thickness. (c) A schematic phase diagram of conductivity for $Sr_2IrO_4$ thin films, where the crossover of conduction mechanism is shown. VRH means variable range hopping, and ES means Efros-Shklovskii. Reproduced with permission from Ref. [90].

*3.3 Tuning via electric current*

Besides the strain engineering, another efficient route to tune the lattice of $Sr_2IrO_4$ is the application of large electric current which was recently addressed by Cao's group [95], an unusual tunability. It was found that the current induced variations along the *a*-axis and *c*-axis are anisotropic, and the *a*-axis expansion can be as large as ~ 1 % if the current reaches a high level [95], as shown in **Figure 7**. More importantly, in accompanying with the huge in-plane lattice expansion, remarkable reductions in both $T_N$ ($\Delta T_N$ ~ 40 K if the current $I$ is as high as ~ 80 mA) and in-plane magnetization ($\Delta M_a$ ~ 0.012 $\mu_B$/Ir at $H$ = 7 T and $I$ = 100 mA) have been observed. The simultaneous modulation of lattice and magnetization can be understood in the framework of pseudospin-lattice locking ($\alpha \sim \phi$). In the meanwhile, other unusual transport behaviors were also observed when the current was huge, including the negative differential resistance and reversible resistance switching [13, 66, 96]. The large current effect demonstrates the potentials for an electro-control of the $J_{\text{eff}}$ = 1/2 state in $Sr_2IrO_4$, although the underlying physics remains elusive.

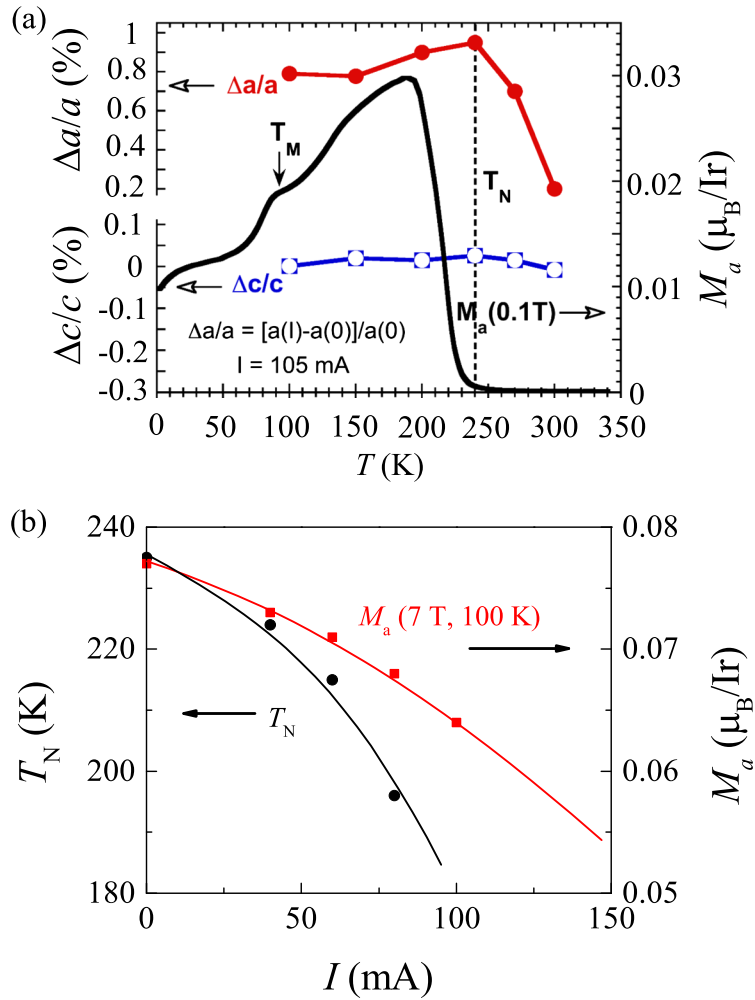

Figure 7. (a) Variations of lattice parameters ($\Delta a/a$ and $\Delta c/c$) as a function of temperature under high electric current $I = 105$ mA, where the $a$-axis shows a large jump at $T_N$ as the $M_a(T)$ curve. $M_a$ is the measured magnetization of the $a$-axis. (b) Measured $T_N$ and $M_a$ (at 7 T and 100 K) as a function of electric current $I$. Reproduced with permission from Ref. [95].

*3.4 Superlattice tuning*

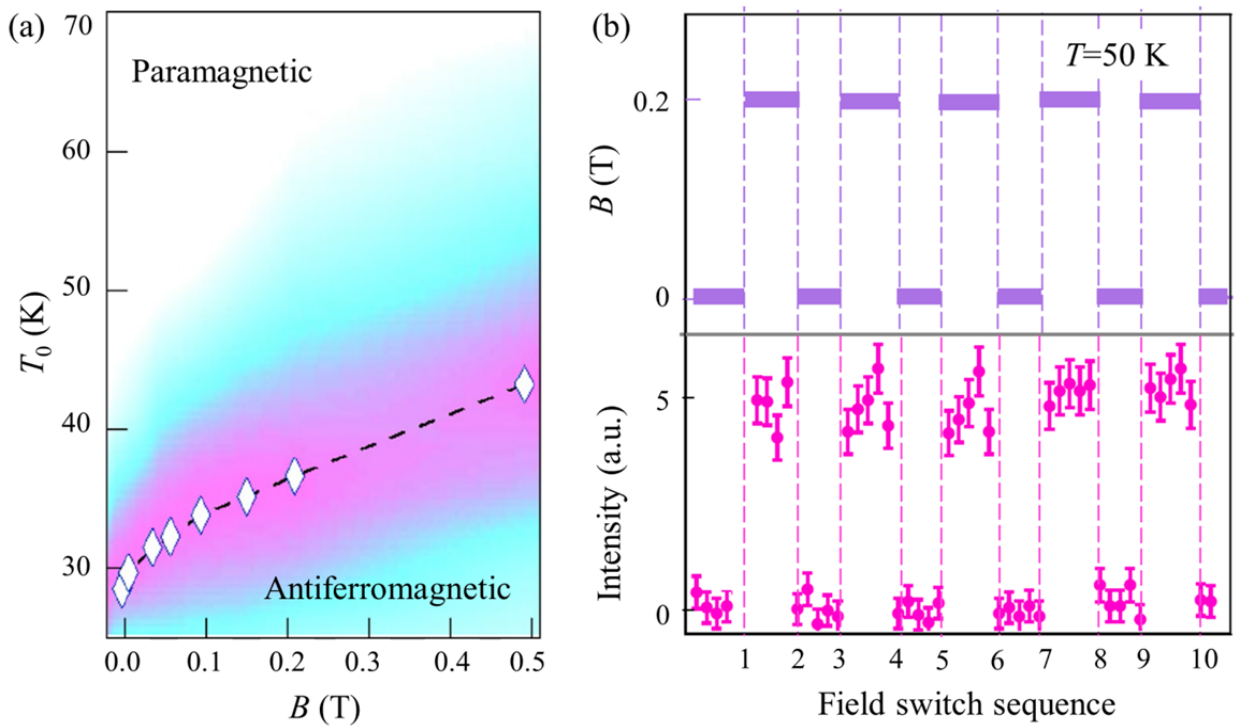

Figure 8. (a) The paramagnetic to antiferromagnetic transition under various magnetic fields, where $T_0$ indicates the crossover temperature. (b) By switching on/off magnetic field $H=0.2$ T at $T = 50$ K, the magnetic peak intensity of the AFM phase shows the corresponding response. Reproduced with permission from Ref. [99].

Finally, one may be noticed that a striking feature of $Sr_2IrO_4$ is the layered magnetic configuration with alternatively stacked $IrO_2$ and SrO layers, in which the cooperation between the interlayer coupling and intralayer interaction determines the magnetic properties. Such layered magnetic structure can be artificially synthesized and engineered, owing to the advance of thin film fabrications. One typical example is the [(SrIrO$_3$)$_n$/(SrTiO$_3$)$_m$] superlattice, where the TiO$_2$ layer is used to replace the IrO$_2$ layer [97-99], so that the IrO$_2$ interlayer distance can be enlarged for weakening the interlayer coupling and suppressing $T_N$. As expected, an increasing thickness of TiO$_2$ layer sandwiched by two IrO$_2$ layers moves $T_N$ down to low-$T$ side significantly, evidencing the pivotal role of interlayer coupling for the long-range AFM order. In the case of $m$ = 2, the superlattice behaves more like a two-dimensional (2D) antiferromagnet which can be effectively controlled by magnetic field $H$ [99]. As shown in **Figure 8**, the AFM state can be switched on and off by a magnetic field as small as $H$ < 0.5 T. Another observation for this case is the weak FM characteristic in the $m$ = 1 superlattice [97], resembling the cases triggered by magnetic field $H$ or Ir-site chemical substitution. It can be concluded that a modulation of the IrO$_2$ interlayer coupling may be crucial for the appearance of weak FM phase. An atomic scale synthesis of the IrO$_2$ layered magnetic structure has been often utilized as an effective tool for cruising fantastic phenomena [100], which nonetheless are non-accessible in Sr$_2$IrO$_4$ and thus will not be described in more details here.

## 4. Giant magnetoresistance due to pseudospin flop

While Sr$_2$IrO$_4$ does possess the robust AFM order, the pseudospin canting still allows small amount of net magnetization, i.e. the weak FM phase with magnetization $M$ up to ~ 0.1 $\mu_B$/Ir [37, 41]. It is the so-called AFM-wFM transition. In fact, Sr$_2$IrO$_4$ has the largest macroscopic magnetization in comparison with other iridates, as we know. The weak FM phase can be stabilized by magnetic field without destructing the inherent AFM order, and this property allows one to access the two degrees of freedom: FM order and AFM order simultaneously, which is unique and broadly interested. A close looking at arrangement of the net magnetic moments $\mu_{net}$ (Fig. 2) suggests a high resembling of the net moment flop transition with the high/low resistance state switching in giant-magnetoresistance (GMR) devices [101, 102]. Every two IrO$_2$ layers with net moments $\mu_{net}$ are separated by a nonmagnetic SrO layer, and neighboring moments $\mu_{net}$ can be switched to either the parallel alignment or the antiparallel one, giving rise to a GMR-like effect. The distinguished point is that this switching is operated in a natural AFM crystal, distinctly different from the mechanism in artificial magnetic multilayers.

Indeed, the resistance ($R$) does show a sudden drop at $H_{flop} \sim 0.2$ T in response to the AFM-wFM transition, as shown in **Figure 9**. The two $R$-states (i.e. high-$R$ state in the low-field region and low-$R$ state in the high-field region), correspond to the antiparallel and parallel alignments of $\mu_{net}$, and the induced MR can be as large as $\sim 80\%$ at low temperature [25]. The observed MR$_c$, namely the out-of-plane MR, as a function of $H$, establishes its one-to-one correspondence with the observed $M(H)$ loop, suggesting the magnetic order origin for the magnetotransport in Sr$_2$IrO$_4$. Notable MR effect can also be seen if current flows within the *ab*-plane (simply called the in-plane MR), but it is clearly smaller than out-of-plane MR [27], as shown in Fig. 9(c) and (d). Such difference between the in-plane and out-of-plane MR data was also observed in traditional GMR structures [103-105], where the difference between the out-of-plane and in-plane MR data was ascribed to the difference in the scaling length [106]. The scaling length of out-of-plane MR is the spin diffusion length, about ten times larger than the scaling length, i.e. the mean free path, of the in-plane MR. The atomic scale GMR-like effect observed in Sr$_2$IrO$_4$ may share similar physics. Nonetheless, we should bear in mind the AFM nature of Sr$_2$IrO$_4$ in terms of the pseudospin alignment and that the magnetic IrO$_2$ layer is atomically thin, different from the artificial GMR heterostructures.

The quest for large MR effect in antiferromagnets has been intensively discussed within the domain of AFM spintronics in the past years [107, 108]. This was initially exploited in AFM heterostructures, i.e. AFM/nonmagnetic-spacer/AFM, in analogy to traditional magnetic multilayers [109, 110]. The AFM counterpart to the GMR, called the AFM-GMR, was theoretically proposed in such AFM structures, and it was found to be purely an interface effect. For instance, a lower resistance will be achieved when the facing layers of antiferromagnets have the same spin orientation [110]. Such scheme has encountered great challenge in experiments, since extremely high quality interface is needed to demonstrate such AFM-GMR effect. This challenge becomes much smoother in natural AFM crystals such as $Sr_2IrO_4$ here that has the natural interfaces. More importantly, the layered AFM structure of $Sr_2IrO_4$ is akin to the AFM multilayers, providing an alternative and more efficient route to obtain significantly large MR. Indeed, a scenario of domain wall scattering was proposed to interpret the MR in $Sr_2IrO_4$ [27], in agreement with the interface scenario in AFM multilayers.

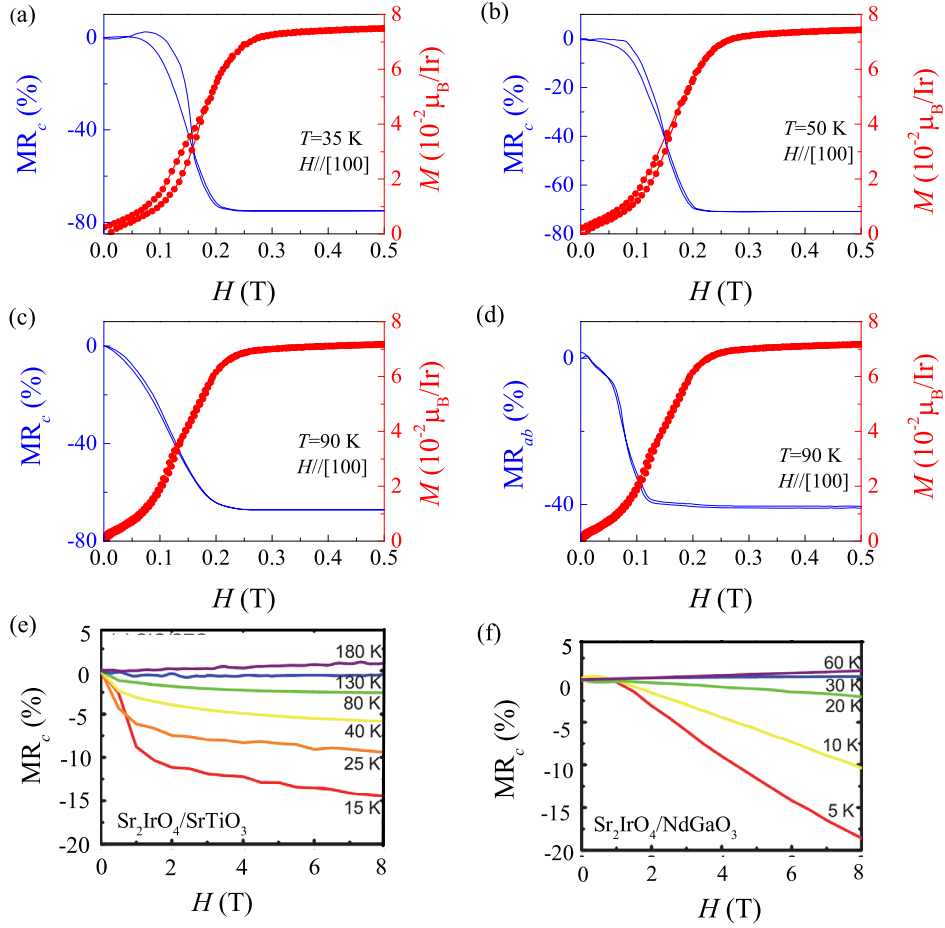

Figure 9. Magnetoresistances (MR$_c$ with current along the *c*-axis and MR$_{ab}$ with current along the *ab*-plane) and in-plane magnetization *M* as a function of *H* measured at: (a) *T* = 35 K, (b) *T* = 50 K, (c) *T* = 90 K. (d) For a comparison, the MR$_{ab}$ data with the in-plane current are also plotted. The MR data of (e) (001) Sr$_2$IrO$_4$/SrTiO$_3$ thin films, and (f) (001) Sr$_2$IrO$_4$/NdGaO$_3$ thin films at different temperatures. Reproduced with permission from Ref. [26].

The atomic scale GMR-like effect, triggered by a field along the *a*-axis (*H*//*a*-axis), can be suppressed quickly by defects, chemical doping, or electrolyte gating [24, 33], since it just originates from the pseudospin flop rather than destruction of the intrinsic AFM order, and thus can be easily destroyed by external perturbations. It is difficult to observe sharp and complete GMR-like effect in Sr$_2$IrO$_4$ thin films in comparison with bulk single crystals even though the thin films show high microstructural quality, owing to SrTiO$_3$ (001) substrate as the best choice to preserve the layered

AFM order and related GMR-like effect in $Sr_2IrO_4$. The largest MR reported so far in $Sr_2IrO_4$/$SrTiO_3$ thin films is ~ 15 % (Fig. 9(e)) and the sudden jump in MR related to the flop transition can still be observable in some experiments [26, 33] but not in others [78]. Large lattice strain can also completely suppress the MR-jump, but instead induce large linear-like MR effect, as identified in $Sr_2IrO_4$/$NdGaO_3$ system (Fig. 9(f)) [26]. Investigation on iridate thin films is still in its infancy (not only for $Sr_2IrO_4$, but for all iridates) and extensive works are desired to fully exploit their electronic properties and emergent functions.

The MR effect driven by the out-of-plane magnetic field ($H$//$c$-axis) also deserves for some discussion, while it looks a bit more complicated. First, whether the flop transition can be triggered by $H$//$c$-axis is still under debates [28, 52]. Second, multiple step-like anomalies in the MR curves were observed in the high field region ($H$ ~ 2 T > $H_{flop}$), but no slop change can be observed in $M(H)$ correspondingly [28]. Therefore, the magnetic scattering scenario may be excluded or it is not the dominant one at least. While the intriguing MR effects for $H$//$c$-axis remain elusive, $H$-driven lattice distortion may be relevant noting the fact that the pseudospin-lattice coupling is essential to describe the magnetism in $Sr_2IrO_4$. Striking magneto-dielectric response was reported in the high field region [111], supporting the argument of strong pseudospin-lattice (phonon) coupling in $Sr_2IrO_4$. Third, the MR magnitudes for the two geometries ($H$//$c$-axis and $H$//$a$-axis) are similar, but the critical fields for pseudospin flop are very different. In addition, so far reported data on the MR data in the $H$//$c$-axis geometry seem to be authors-dependent, an issue to be clarified [27, 28].

## 5. Anisotropic magnetoresistance toward antiferromagnetronics

The atomic scale GMR-like effect has illustrated the uniqueness of $Sr_2IrO_4$ as an AFMtronics candidate. More than this, a system with a novel $J_{eff} = 1/2$ state does host more exciting functionalities, including remarkable and controllable anisotropic magnetoresistance (AMR) and nonvolatile memory [23, 25]. The AMR effect is the magnetotransport counterpart of the relativistic energy anisotropy, a powerful and widely utilized function for detecting reorientation of magnetic moment typically in FM materials. Recently, this effect was demonstrated in some AFM materials [112-114] since it is an even function of magnetization. A seminal work on this topic is the observation of large tunneling AMR in IrMn-based AFM tunnel junctions [115], and the essential role of large SOC in inducing such AFM-based AMR (AFM-AMR) effect has been emphasized.

Along this line, it was reported that the $Sr_2IrO_4/La_{2/3}Sr_{1/3}MnO_3$ heterostructure indeed exhibits AMR effect up to ~1 % [21], where the FM $La_{2/3}Sr_{1/3}MnO_3$ layer was used to control the AFM lattice of $Sr_2IrO_4$ through the interface coupling. Subsequent experiments with $Sr_2IrO_4$ bulk crystals found that FM buffer layer is not necessary to generate AMR, because $Sr_2IrO_4$ has the nonzero net magnetization at $H > H_{flop}$ [22, 24]. It is believed that realizing the AMR effect in a pure AFM material without any auxiliary reference layer is an important step by which the merit of AFM spintronic properties such as ultrafast spin dynamics in THz domain can be truly taken.

The AFM-AMR was also identified in $Sr_2IrO_4$ thin films grown on (001) $SrTiO_3$ substrates without reference layer [23, 26, 33]. Because of the good lattice fit between $Sr_2IrO_4$ and $SrTiO_3$, it is allowed to preserve the spin-orbit coupled AFM order in the epitaxial thin films. The MR curves with $H$ applied along the [100] (easy axis) and the [110] (hard axis) directions present an intriguing intercross in the high field region, indicating unconventional AMR in the films [23]. This was further confirmed by the detailed $R(\Phi)$ measurements under various $H$, as shown in **Figure 10**. At low field, the fourfold crystalline AMR with minima at the easy axes is presented. Upon increasing $H$, the AMR contour is rotated by ~ 45°, indicating that the pristine fourfold AMR minima are now changed to the maximal positions in the induced fourfold AMR. This AMR contour rotation can be even observed in the high-$T$ range close to $T_\mathrm{N}$~240 K. The first-principles calculations revealed the different charge-gaps as the pseudospins point along the easy- and hard-axes. For instance, a larger charge-gap is expected when the pseudospins are aligned along the easy axis, in comparison with the case of pseudospins along the hard axis. This is consistent with the $J_\mathrm{eff}= 1/2$ picture in which the strong SOC is essentially involved in developing electronic structure of $Sr_2IrO_4$.

Similar AMR contour rotation was also observed in $Sr_2IrO_4$ bulk crystals while the critical field is smaller. Surprisingly, a giant AMR ratio reaching ~ 160 % was identified in bulk $Sr_2IrO_4$, which is a record in the field of AFMtronics [25]. The MR curves show sudden change at low field associated with the pseudospin flop transition, as discussed in Sec. 4. As shown in **Figure 11**, the critical fields for the MR-jumps show clear difference when $H$ is applied along the easy axis (the [100] direction) and in-plane hard axis (the [110] direction), leading to a window in this region. The difference in measured $R$ for the two cases can be as large as $\Delta R$ ~ 38 kΩ at $T$ = 35 K, a value larger than the most reported values in literature, including those in semiconductors and tunnel junctions which usually show relatively large $\Delta R$ and AMR. By plotting the AMR ratio as a function of $H$, one could easily see that the remarkable AMR enhancement is tightly related to the flop transition.

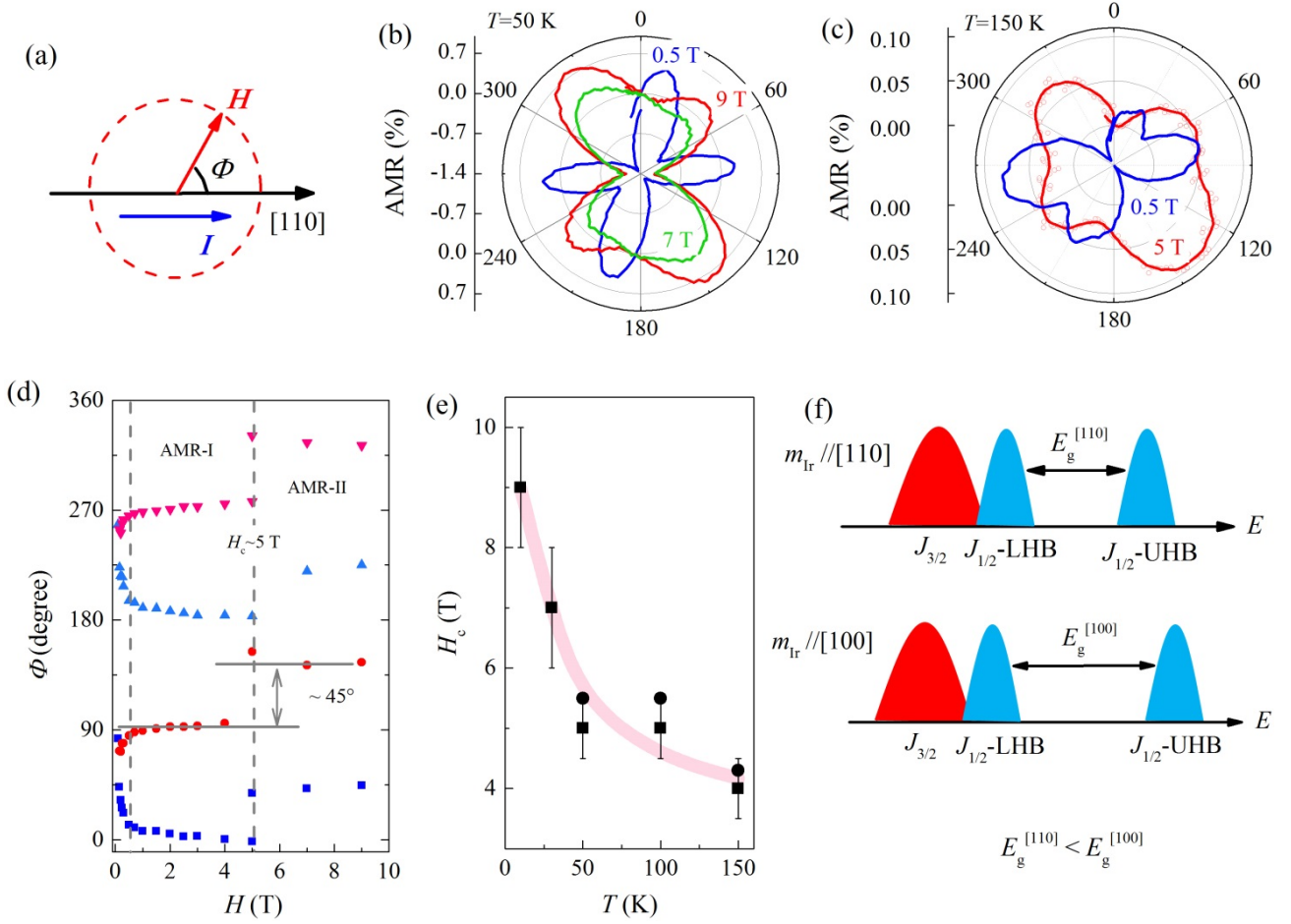

Figure 10. (a) The geometry for AMR measurements, where $\Phi$ is the angle between $H$ and the [110] direction. The AMR angle-dependent contours at several $H$ for (b) $T = 50$ K, and (c) $T = 150$ K. (d) The peak positions of AMR curves as a function of $H$. (e) The critical field $H_c$ (squares) for the AMR contour switching as a function of temperature. Dots are the fields where MR curves of $H$//[100] and $H$//[110] show intercross. (f) A sketch of band structure for the cases of $m_{Ir}$//[100] and $m_{Ir}$//[110] where $m_{Ir}$ is the pseudospin moment of Ir. Reproduced with permission from Ref. [23].

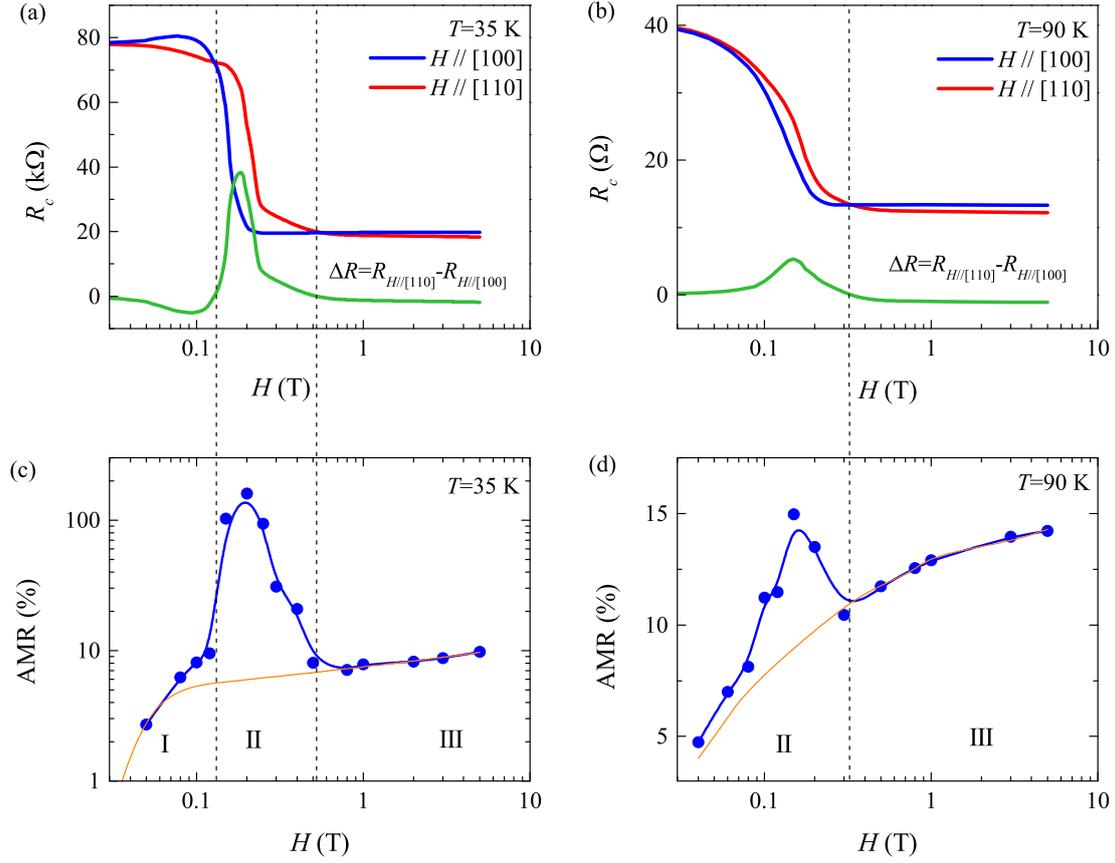

Figure 11. The measured resistance along the *c*-axis, $R_c$, as function of $H$ at (a) $T$ = 35 K, and (b) $T$ = 90 K. Green curves are the estimated difference in $R_c$ with $H$ along different directions. The estimated AMR magnitude as a function of $H$ at (c) $T$ = 35 K, and (d) $T$ = 90 K. Orange lines are used to highlight the remarkable enhancement in AMR. Reproduced with permission from Ref. [25].

There are two factors pivotal for the observation of large AMR in $Sr_2IrO_4$. One is that the flop transition has to be sharp and complete, which promises a giant MR. This looks sensitive to external perturbations such as chemical doping and strain in thin films. The other one is the in-plane magnetic anisotropy, allowing us to have a window between the MR curves with $H$//[100] and $H$//[110]. The in-plane magnetic anisotropy of $Sr_2IrO_4$ has been a missed issue until very recently. In previous studies focusing on the exploration of AFM-AMR, magnetocrystalline anisotropy has been considered solely to realize AMR in antiferromagnets, and the observed AMR ratio is unfortunately small overall [107, 108, 116]. Therefore, the observation of giant AMR in $Sr_2IrO_4$ has illustrated an efficient mean, i.e. combination of various mechanisms, to enhance the AFM-AMR remarkably.

By tiny Ga-doping (1.0 %) at Ir-site in $Sr_2IrO_4$, nonvolatile memory was demonstrated, and the resistive ratio of two memory states can be as large as 4.5 % [25], as shown in **Figure 12**. The AFM-based memory phenomena have been identified in several antiferromagnets such as MnTe and FeRh, and a heat assist magnetic recording method was generally used to manipulate the memory states [113, 114]. For instance, in order to set the memory states in FeRh which represents a rare room temperature AFM memory resistor, the sample has to be warmed up above $T_N$, and then cooled down with $H$. In $Sr_2Ir_{0.99}Ga_{0.01}O_4$, the memory states can be switched *in-situ* by changing the direction of $H$, and the memory effect is nonvolatile and fully reproducible in the successive write-read cycles, which looks more compatible with realistic devices.

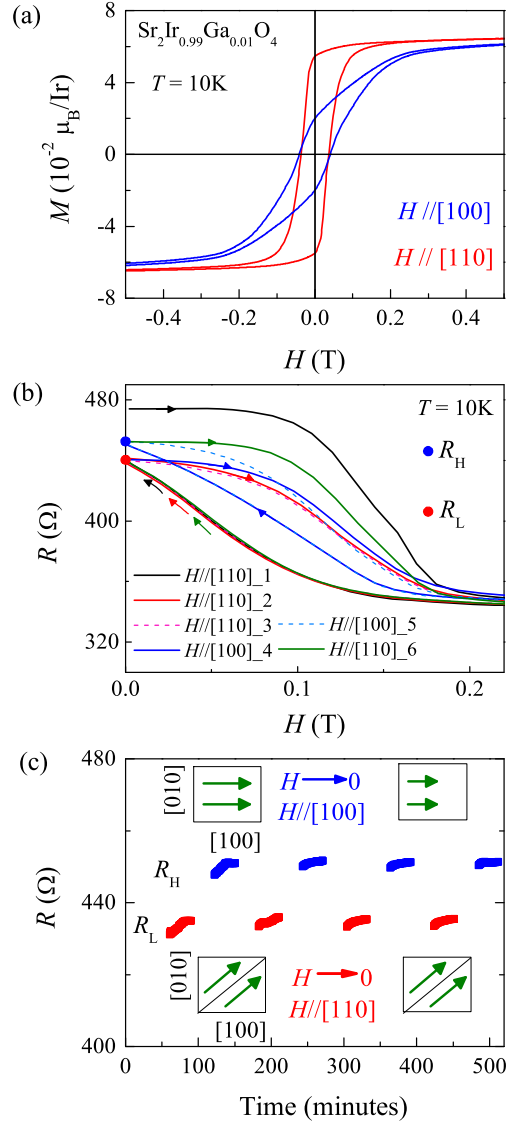

Figure 12. The measured physical properties for $Sr_2Ir_{0.99}Ga_{0.01}O_4$ crystals at $T$ = 10 K. (a) Magnetization $M$ as a function of $H$. (b) Resistance $R$ as a function of $H$ under various $H$ cycling procedures. (c) The $H$-switching of $R$ between different nonvolatile memory states. Reproduced with permission from Ref. [25].

Increasing Ga-doping content in $Sr_2Ir_{1-x}Ga_xO_4$ induced metallic transport behavior, while the canted AFM order was maintained just showing gentle decrease in $T_N$ from 240 K to 180 K [24]. Note that it seems a bit difficult to successively tune electric transport from insulating to metallic without breaking the pristine magnetic order in AFM materials. Upon Ga-doping, the fourfold crystal AMR symmetry can be well preserved, although its magnitude is reduced from ~ 16 % to 1.0 %, as shown in **Figure 13**. The suppression of AMR magnitude is much more obvious when samples host insulating transport. In the samples with $x > 0.05$ showing metallic behavior, the AMR magnitude evolves with $x$ steadily at a level of ~ 1.0 % at low temperature. This is in fact a relatively large AMR ratio in comparison with other metallic AFM materials.

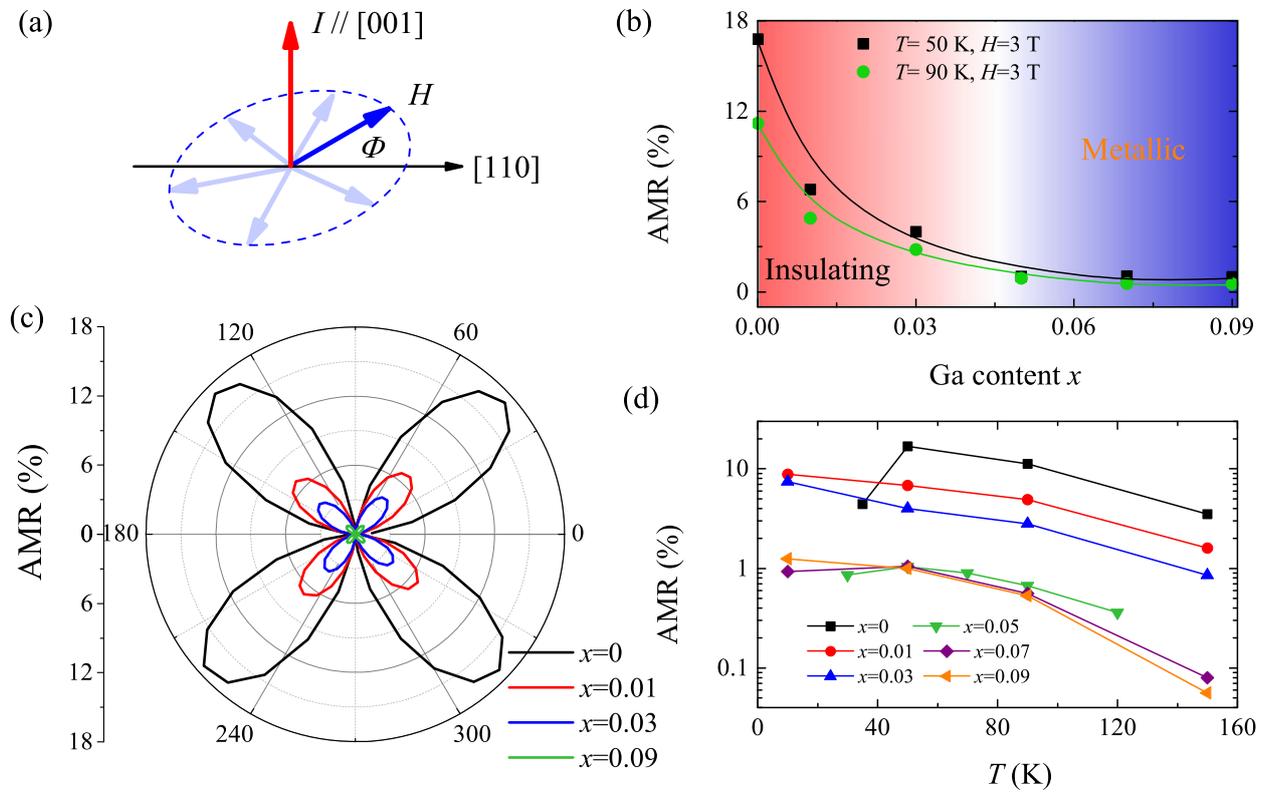

Figure 13. (a) A sketch of the AMR measurements for $Sr_2Ir_{1-x}Ga_xO_4$ crystals ($0 \leq x \leq 0.09$), where the current is applied along the *c*-axis and $H$ is rotated within the *ab*-plane. (b) The AMR magnitude as a function of Ga-doping content. (c) The fourfold AMR symmetry is observed for all samples with different Ga-content. (d) The AMR ratio as a function of $T$ for different $Sr_2Ir_{1-x}Ga_xO_4$ crystals. Adapted with permission from Ref. [24].

## 6. Final remarks and outlook

To this end, we have presented a brief overview on the recent progress of $Sr_2IrO_4$ with novel $J_{eff}$ = 1/2 state, addressing the magnetic ground state, fantastic electronic properties, canted antiferromagnetism and related magnetotransport behaviors. $Sr_2IrO_4$ represents a leading iridate that has been intensively addressed recent years. Although the predicted unconventional superconductivity has not yet been observed, a bunch of fascinating phenomena have been identified, including low energy excitations, Fermi arcs, pseudogap, electronic inhomogeneity, and broken symmetry phases. In the meanwhile, our understanding of the quasi 2D antiferromagnetism associated with the $J_{eff}$ = 1/2 moments has been advanced largely. Apart from the metal-insulator transition, the highly tunable layered AFM phase and associated phenomena such as spin-density wave, time-reversal symmetry breaking hidden order, and pseudospin flop transitions have been identified. Especially, the observations of atomic scale GMR-like effect, remarkable and controllable AMR, and nonvolatile memory have highlighted $Sr_2IrO_4$ as a promising AFMtronic candidate.

While these observations on one hand underscore the uniqueness of $Sr_2IrO_4$ in both physical and functional aspects, a large array of focusing issues remain elusive on the other hand. One of the most attractive topics is experimentally yet missed unconventional superconductivity in carrier-doped $Sr_2IrO_4$. Although many strategies have been proposed, such as pushing the hidden phase into its quantum critical point, increasing chemical doping concentration effectively, efficient means to approach the elusive superconducting state remain yet unexploited. This is obviously a very meaningful issue, and extensive works are required to fill the huge gap between the related theories and experiments. Regarding the canted AFM phase, it can now be more precisely described using a modified Heisenberg Hamiltonian. However, after involving stimuli such as dopants, strain, and electric field, things are going complex and no certain tendency and principle can be concluded. For instance, in some cases, the long-range order collapses simultaneously when the Mott gap is closed, indicating the close correlation between charge and magnetic degrees of freedom. This is consistent with the close energy scale between the exchange (~ 0.1 eV) and charge gap (~ 0.5 eV). However, there are still a few of experimental results showing the decoupled long-range order and insulating state.

The demonstration of AFMtronic functionalities made $Sr_2IrO_4$ even more tantalizing, distinguishing this material from other iridates. Obviously, the magnitude of MR and AMR are already sufficient large. However, these superior properties are more frequently observed in low temperature range. For all iridates that have been discovered, one may fall frustrated indeed, since the highest reported Néel point is just ~ 280 K in $Sr_3Ir_2O_7$. Nevertheless, it has been reported that $T_N$ can be enhanced significantly by simply enhancing the interlayer coupling. Another interesting phenomenon is the nonvolatile memory in $Sr_2IrO_4$ with 1% Ga-doping. Again, the operating temperature is low. Although these works have shown the fascinating aspects of $Sr_2IrO_4$, there is much room for utilizing these functionalities in practical devices.

Thin film fabrication is a key step to integrate this leading iridate, $Sr_2IrO_4$, into practical devices, and it has been revealed that the electronic properties of $Sr_2IrO_4$ can be efficiently tuned through strain indeed. However, related experiments are largely unexplored. One possible reason is the lack of high quality samples. High quality pure phase $Sr_2IrO_4$ can only be stabilized within a narrow growth window deposition. In fact, even for $Sr_2IrO_4$ bulk crystals, the results from different groups can be somehow scattering. For instance, the step-anomaly appearing in the $M(H)$ curves with $H//c$-axis is not yet understood. In some cases, strong metallicity can arise in La-doped $Sr_2IrO_4$.

However, qualitatively different electric transport behavior with clear upturn in the $R(T)$ curve at low temperature has also been reported. These inconsistent results suggest the important role of sample quality, which has been much less focused but surely deserves special attention.

**ACKNOWLEDGEMENTS**

This work is supported by the National Nature Science Foundation of China (Grant Nos. 11774106, 51431006, and 51721001), the National Key Research Projects of China (Grant No. 2016YFA0300101), and the Fundamental Research Funds for the Central Universities (Grant No. 2019kfyRCPY081, 2019kfyXKJC010).

Received: ((will be filled in by the editorial staff))
Revised: ((will be filled in by the editorial staff))
Published online: ((will be filled in by the editorial staff))

**Table of Content:**

Perovskite $Sr_2IrO_4$ hosts a novel $J_{eff} = 1/2$ state related to strong spin-orbit coupling, and represents a rare example of iridates that has been better understood both theoretically and experimentally. In the last decade, fascinating electronic properties such as possible unconventional superconductivity and giant magnetoresistance were revealed in this material. This progress report overviews these developments in two aspects: materials fundamentals and functionality potentials.

**Keywords:** iridates, spin-orbit coupling, $J_{eff} = 1/2$ state, antiferromagnetic, anisotropic magnetoresistance.

Chengliang Lu* and Jun-Ming Liu*

**The $J_{eff}$=1/2 antiferromagnet $Sr_2IrO_4$: A golden avenue toward new physics and functions**

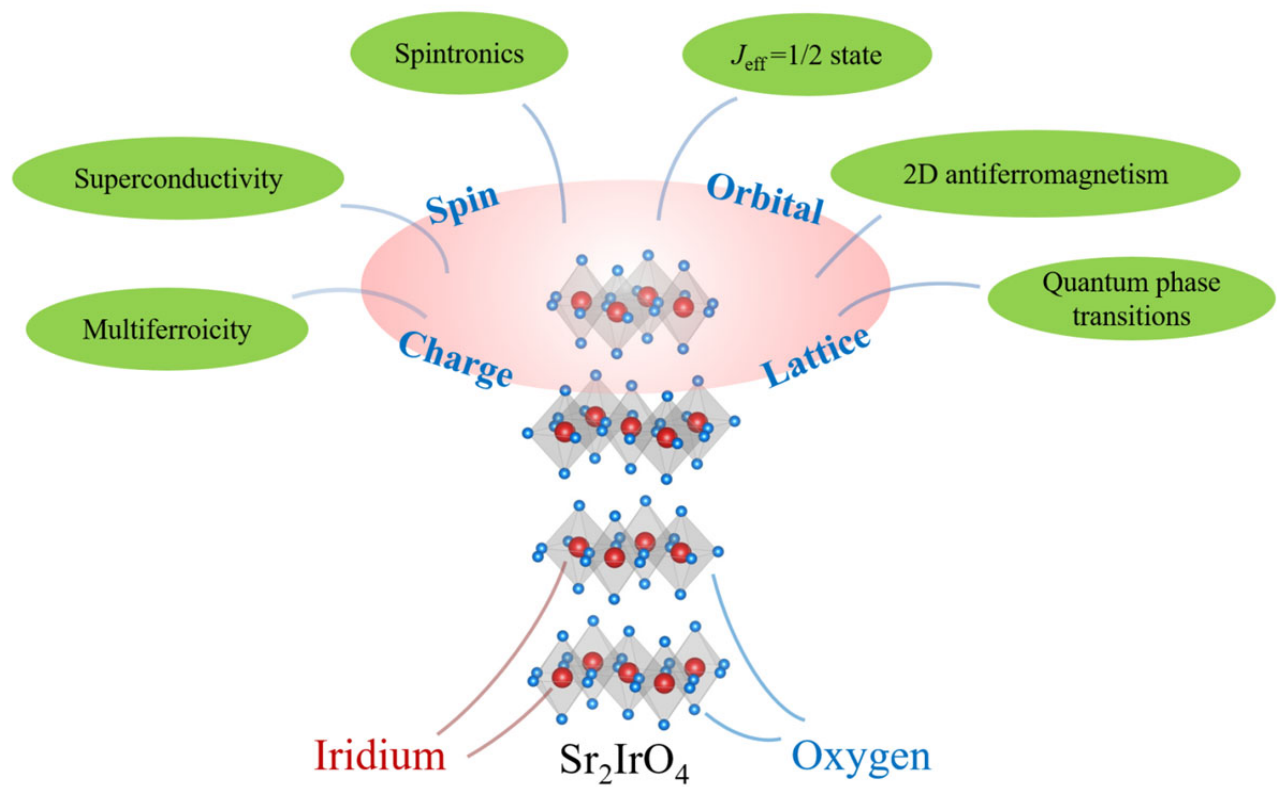